# Propositions in Linear Multirole Logic as Multiparty Session Types

Hongwei Xi Hanwen Wu
Boston University
{hwxi,hwwu}@cs.bu.edu

## Abstract

We identify multirole logic as a new form of logic and formalize linear multirole logic (LMRL) as a natural generalization of classical linear logic (CLL). Among various meta-properties established for LMRL, we obtain one named multi-cut elimination stating that every cut between three (or more) sequents (as a generalization of a cut between two sequents) can be eliminated, thus extending the celebrated result of cut-elimination by Gentzen. We also present a variant of $\pi$-calculus for multiparty sessions that demonstrates a tight correspondence between process communication in this variant and multi-cut elimination in LMRL, thus extending some recent results by Caires and Pfenning (2010) and Wadler (2012), among others, along a similar line of work.

## 1. Introduction

While the notion of multirole logic stems directly from studies on multiparty sessions (Honda et al. 2008), we see it beneficial to start with dyadic sessions (Honda 1993; Takeuchi et al. 1994). In broad terms, a dyadic session is an interaction between two concurrently running programs, and a session type is a form of type for specifying sessions. As an example, let us assume that two programs P and Q are connected with a bidirectional channel (that is, a channel with two endpoints). From the perspective of P, the channel may be specified by a term sequence of the following form:

$$\texttt{snd}(\textbf{int}) :: \texttt{snd}(\textbf{int}) :: \texttt{rcv}(\textbf{bool}) :: \texttt{nil}$$

which means that an integer is to be sent, another integer is to be sent, a boolean is to be received, and finally the channel is to be closed. Clearly, from the perspective of Q, the channel should be specified by the following term sequence:

$$\texttt{rcv}(\textbf{int}) :: \texttt{rcv}(\textbf{int}) :: \texttt{snd}(\textbf{bool}) :: \texttt{nil}$$

which means precisely the dual of what the previous term sequence does. We may think of P as a client who sends two integers to the server Q and then receives from Q either true or false depending on whether or not the first sent integer is less than the second one.

A simple but crucial observation is that the above two term sequences can be unified as follows:

$$\texttt{msg}(0, 1, \textbf{int}) :: \texttt{msg}(0, 1, \textbf{int}) :: \texttt{msg}(1, 0, \textbf{bool}) :: \texttt{nil}$$

```
fun P() = let
  val () =
    channel_send(CH0, I1, 0, 1) // send to Q
  val () =
    channel_send(CH0, I2, 0, 1) // send to Q
  val b0 = channel_recv(CH0, 1, 0) // recv from Q
  val () = channel_close(CH0) // close the P-end of CH
in b0 end (* end of [P] *)

fun Q() = let
  val i1 =
    channel_recv(CH1, 0, 1) // recv from P
  val i2 =
    channel_recv(CH1, 0, 1) // recv from P
  val () =
    channel_send(CH1, i1 < i2, 1, 0) // send to P
  val () = channel_close(CH1) // close the Q-end of CH
in () end (* end of [Q] *)
```

**Figure 1.** Some pseudo code in ML-like syntax

where 0 (client) and 1 (server) refer to the two roles implemented by P and Q, respectively. Given a type $T$ and two roles $i$ and $j$, the term $\texttt{msg}(i, j, T)$ basically indicates a value of the type $T$ being transferred from a party implementing role $i$ to another party implementing role $j$. In particular, $\texttt{msg}(i, j, T)$ is interpreted as a send (receive) operation by a party implementing role $i$ ($j$).

In Figure 1, we present some pseudo code showing a plausible way to implement the programs P and Q. Please note that the functions P and Q, though written together here, can be written in separate contexts. We use CH0 and CH1 for the two endpoints of some channel CH (assumed to be available in the surrounding context of the code) and I1 and I2 for two integers; the functions `channel_send` and `channel_recv` are for sending and receiving data via a given channel (endpoint), and `channel_close` for closing one. In the following presentation, we use the name *full channel* for a channel (like CH) and instead refer to each endpoint of a full channel as a channel. For instance, CH0 (CH1) is a channel of role 0 (1), which is held by a party implementing role 0 (1).

Let us sketch a way to make the above pseudo code type-check. Given an integer $i$ and a session type $S$, let $\textbf{chan}(i, S)$ be the type for a channel of role $i$. We can assign the following type to `channel_send`:

$$(!\textbf{chan}(i, \texttt{msg}(i, j, T) :: S) \gg \textbf{chan}(i, S), \textbf{int}(i), \textbf{int}(j), T) \rightarrow \textbf{1}$$

where $i \neq j$ is assumed to hold, and $\textbf{int}(i)$ and $\textbf{int}(j)$ are singleton types for integers equal to $i$ and $j$, respectively, and $T$ and $S$ stand for a type and a session type, respectively. Basically, this type[1] means that calling `channel_send` on a channel of the type

[1] Strictly speaking, this type should be referred to as a type schema as it contains occurrences of meta-variables.

$\textbf{chan}(i, \mathtt{msg}(i, j, T) :: S)$, integer $i$, integer $j$ and a value of the type $T$ returns a unit while *changing* the type of the channel to $\textbf{chan}(i, S)$. Clearly, **chan** is required to be a linear type constructor for this to make sense. As can be expected, the type assigned to `channel_recv` should be of the following form:

$$(!\textbf{chan}(j, \mathtt{msg}(i, j, T) :: S) \gg \textbf{chan}(j, S), \textbf{int}(i), \textbf{int}(j)) \to T$$

where $i \neq j$ is assumed to hold. This type indicates that calling `channel_recv` on a channel of the type $\textbf{chan}(j, \mathtt{msg}(i, j, T) :: S)$, integer $i$ and integer $j$ returns a value of the type $T$ while *changing* the type of the channel to $\textbf{chan}(j, S)$. As for `channel_close`, it is assigned the following type:

$$(\textbf{chan}(i, \mathtt{nil})) \to \mathbf{1}$$

indicating that calling `channel_close` on a channel consumes the channel (so that the channel is no longer available for use).

Assume that $\mathrm{CH}^a$ and $\mathrm{CH}^b$ are two full channels specified by a session type $S$. The two endpoints $\mathrm{CH}^a_0$ and $\mathrm{CH}^a_1$ of $\mathrm{CH}^a$ are given the types $\textbf{chan}(0, S)$ and $\textbf{chan}(1, S)$, respectively. Similarly, the two endpoints $\mathrm{CH}^b_0$ and $\mathrm{CH}^b_1$ of $\mathrm{CH}^b$ are given the types $\textbf{chan}(0, S)$ and $\textbf{chan}(1, S)$, respectively. If some party holds both $\mathrm{CH}^a_1$ and $\mathrm{CH}^b_0$, then the party can link them together by performing a form of a bidirectional forwarding that sends onto $\mathrm{CH}^b_0$ each value received on $\mathrm{CH}^a_1$ and vice versa. After $\mathrm{CH}^a_1$ and $\mathrm{CH}^b_0$ are linked in such a manner, $\mathrm{CH}^a_0$ and $\mathrm{CH}^b_1$ can be seen as the tw endpoints of a full channel specified by $S$. It is well-known that bidirectional forwarding between two matching channels (of types $\textbf{chan}(0, S)$ and $\textbf{chan}(1, S)$ for some $S$) corresponds to cut-elimination in linear logic (Caires and Pfenning 2010; Wadler 2012a).

Instead of two roles, let us assume the availability of three roles 0, 1 and 2. One may be tempted to guess that the aforementioned bidirectional forwarding between two channels can be generalized to work in the case of three channels of types $\textbf{chan}(0, S)$, $\textbf{chan}(1, S)$, and $\textbf{chan}(2, S)$ for some $S$. Assume that $\mathrm{CH}^a$, $\mathrm{CH}^b$ and $\mathrm{CH}^c$ are three full channels specified by a session type $S$. For each $x \in \{a, b, c\}$ and $i \in \{0, 1, 2\}$, $\mathrm{CH}^x_i$ (as an endpoint of $\mathrm{CH}^x$) is of the type $\textbf{chan}(i, S)$. If $\mathrm{CH}^a_2$, $\mathrm{CH}^b_1$ and $\mathrm{CH}^c_0$ are chosen to be linked together so that each value received on one of them is sent onto another of them, then the other 6 endpoints $\mathrm{CH}^a_0$,$\mathrm{CH}^a_1$, $\mathrm{CH}^b_0$,$\mathrm{CH}^b_2$, $\mathrm{CH}^c_1$ and $\mathrm{CH}^a_2$ should form another full channel. This is certainly unexpected (if not unsound) as each full channel is assumed to have only three endpoints: one for each of the three roles 0, 1 and 2. As a consequence, we introduce multirole channels as follows.

Given a role $i$ and a session type $S$, the type $\textbf{chan}(i, S)$ for single-role channels can be naturally transitioned into one of the form $\textbf{chan}(R, S)$ for multirole channels, where $R$ stands for a set of roles. In particular, $\textbf{chan}(i, S)$ can be simply treated as $\textbf{chan}(\{i\}, S)$. For notational convenience, we may simply write $i$ for $\{i\}$ from this point on. Assume that there exists a fixed set of $N$ roles ranging from 0 to $N-1$ for some natural number $N \geq 2$. For each $R$, we use $\overline{R}$ for the complement of $R$, which consists of all of the natural numbers less than $N$ that are not in $R$. In particular, $\overline{\emptyset}$ refers to the set $\{0, 1, \ldots, N\}$. Each full channel CH specified by $S$ may have $n$ endpoints $\mathrm{CH}_{R_i}$ that are assigned the types $\textbf{chan}(R_i, S)$ for $i = 1, \ldots, n$, where $R_1, \ldots, R_n$ form a partition of $\overline{\emptyset}$. If a value is sent onto one of the endpoints, then this value is supposed to reach all of the other endpoints. In other words, sending simply acts like broadcasting.

We may refer to a channel as a channel of roles $R$ if the channel is assigned a type of the form $\textbf{chan}(R, S)$. We have the following two scenarios for interpreting $\mathtt{msg}(i, j, T)$ based on a given set $R$:

- Assume $i \in R$. Then any party holding a channel of type $\textbf{chan}(R, \mathtt{msg}(i, j, T) :: S)$ is supposed to send onto the channel a tagged value in which the tag is $j$ and the value is of type $T$. As the channel is an endpoint of a full channel, the tagged value should reach all of the other endpoints of the full channel.
- Assume $i \notin R$. Then any party holding a channel of type $\textbf{chan}(R, \mathtt{msg}(i, j, T) :: S)$ is supposed to receive on the channel a tagged value in which the tag is $j$ and the value is of type $T$.

We may stipulate that any party should discard a tagged value received on a channel of roles $R$ if the attached tag does not belong to $R$. If broadcasting from one endpoint of a full channel to the others are built on top of point-to-point communication[2], then this stipulation implies no need for actually sending a tagged value to a channel of roles $R$ whenever the attached tag does not belong to $R$. With the stipulation, we have the following four scenarios for interpreting $\mathtt{msg}(i, j, T)$ based on a given set $R$ of roles:

- Assume $i \in R$ and $j \in R$. Then any party holding a channel of type $\textbf{chan}(R, \mathtt{msg}(i, j, T) :: S)$ should ignore the term $\mathtt{msg}(i, j, T)$ as there is no other endpoint expecting to receive a value tagged with $j$.
- Assume $i \notin R$ and $j \notin R$. Then any party holding a channel of type $\textbf{chan}(R, \mathtt{msg}(i, j, T) :: S)$ should ignore the term $\mathtt{msg}(i, j, T)$ as it is expected to neither send nor receive.
- Assume $i \in R$ and $j \notin R$. Then any party holding a channel of type $\textbf{chan}(R, \mathtt{msg}(i, j, T) :: S)$ should send a value of type $T$ (to the only other endpoint expecting to receive such a value).
- Assume $i \notin R$ and $j \in R$. Then any party holding a channel of type $\textbf{chan}(R, \mathtt{msg}(i, j, T) :: S)$ should receive a value of type $T$ (from the only other endpoint expecting to send such a value).

With the above interpretation, `channel_send` can be assigned the following type:

$$(!\textbf{chan}(R, \mathtt{msg}(i, j, T) :: S) \gg \textbf{chan}(R, S), \textbf{int}(i), \textbf{int}(j), T) \to \mathbf{1}$$

where $i \in R$ and $j \notin R$ is assumed; `channel_recv` can be assigned the following type:

$$(!\textbf{chan}(R, \mathtt{msg}(i, j, T) :: S) \gg \textbf{chan}(R, S), \textbf{int}(i), \textbf{int}(j)) \to T$$

where $i \notin R$ and $j \in R$ is assumed. As for `channel_close`, the following type is assigned:

$$(\textbf{chan}(R, \mathtt{nil})) \to \mathbf{1}$$

In addition, we need to introduce a function `channel_skip` of the following type:

$$(!\textbf{chan}(R, \mathtt{msg}(i, j, T) :: S) \gg \textbf{chan}(R, S)) \to \mathbf{1}$$

where either $i, j \in R$ or $i, j \notin R$ is assumed. Note that `channel_skip` is really a proof function in the sense that it does nothing at runrime.

In the case where $\overline{\emptyset} = \{0, 1\}$, a party holding two channels of types $\textbf{chan}(0, S)$ and $\textbf{chan}(1, S)$ can link them together by performing bidirectional forwarding (of values received on them) if each channel is an endpoint of a distinct full channel. In the general case where $\overline{\emptyset}$ may contain more than 2 roles, a party holding $n$ channels of types $\textbf{chan}(R_i, S)$ for $i = 1, \ldots, n$ can link them together if each channel is an endpoint of a distinct full channel and the role sets $\overline{R}_1, \ldots, \overline{R}_n$ form a partition of $\overline{\emptyset}$, that is, the following equality holds:

$$\overline{R}_1 \uplus \ldots \uplus \overline{R}_n = \overline{\emptyset}$$

where $\uplus$ refers to the union of two disjoint sets. For instance, we may have $R_1 = \{0, 1\}$, $R_2 = \{0, 2\}$, and $R_3 = \{1, 2\}$ in the case where $\overline{\emptyset} = \{0, 1, 2\}$. The actual linking of such $n$ channels can be performed as follows:

[2] For instance, this is the case in an experimental implementation we did where the underlying communication is based on shared memory.

- Assume that a tagged value is received on one channel. Then a type of the form $\mathbf{chan}(R_i, \mathtt{msg}(p,q,T) :: S)$ is assigned to the channel originally, where $p \notin R_i$ holds. Clearly, we have $p \in \overline{R}_i$ and thus $p \in R_j$ for any role $j$ that is not $i$ (since $\overline{R}_1, \ldots, \overline{R}_n$ form a partition of $\overline{\emptyset}$), which means that the received tagged value can be sent onto each of the remaining $n-1$ channels.

It can be readily verified that the $n$ distinct full channels involved in such an act of linking form another full channel at the end. As an example, let us assume the existence of 3 full channels $\mathrm{CH}^a$, $\mathrm{CH}^b$ and $\mathrm{CH}^c$ that are all specified by $S$. For $\mathrm{CH}^a$, there are two endpoints $\mathrm{CH}^a_0$ and $\mathrm{CH}^a_{0'}$ that are of types $\mathbf{chan}(0, S)$ and $\mathbf{chan}(0', S)$, respectively, where 0 refers to the singleton set $\{0\}$ and $0'$ the complement of $\{0\}$. For $\mathrm{CH}^b$, there are two endpoints $\mathrm{CH}^b_1$ and $\mathrm{CH}^b_{1'}$ that are of types $\mathbf{chan}(1, S)$ and $\mathbf{chan}(1', S)$, respectively. For $\mathrm{CH}^c$, there are two endpoints $\mathrm{CH}^c_2$ and $\mathrm{CH}^c_{2'}$ that are of types $\mathbf{chan}(2, S)$ and $\mathbf{chan}(2', S)$, respectively. The aforementioned act of linking can be performed on the 3 endpoints $\mathrm{CH}^a_{0'}$, $\mathrm{CH}^b_{1'}$ and $\mathrm{CH}^c_{2'}$, resulting in the formation of a full channel consisting of the other 3 endpoints $\mathrm{CH}^a_0$, $\mathrm{CH}^b_1$ and $\mathrm{CH}^c_3$.

While we made use of some linearly typed functions on channels to illustrate the notion of multirole, we do not attempt to formally study such functions in this paper. Instead, we focus on logic. Since the act of linking two matching channels can be given an interpretation based on cut-elimination in intuitionistic linear logic (Caires and Pfenning 2010) and classical linear logic (Wadler 2012a), we naturally expect that the act of linking $n$ matching channels (for $n \geq 2$) can be interpreted similarly based on cut-elimination in a linear logic of certain kind. We are able to form linear multirole logic (LMRL) to serve this purpose precisely. For long, studies on logics have been greatly influencing research on programming languages. In the case of LMRL, we see a genuine example that demonstrates the influence of the latter on the former.

The rest of the papers is organized as follows. We formulate LMRL in Section 2, establishing various meta-properties for them. We primarily focus on conjunctive LMRL ($\mathrm{LMRL}_\wedge$) while briefly mentioning disjunctive LMRL ($\mathrm{LMRL}_\vee$) as the dual of *LMRLconj*. We then present in Section 3 a process calculus $\pi$LMRL, which can be seen as a typed variant of $\pi$-calculus. The session types in $\pi$LMRL are just the formulas in LMRL and the reduction semantics of $\pi$LMRL is directly based on cut-elimination in LMRL. Lastly, we mention some closely related and conclude.

The primary contribution of the paper lies in the identification of multirole logic as a new form of logic and the presented formalization of linear multirole logic (LMRL). We consider the formulation and proof of various meta-properties on LMRL a large part of this contribution. In particular, we formulate a cut-rule for multiple sequents in LMRL and prove its admissibility, naturally extending the celebrated result of cut-elimination by Gentzen (Gentzen 1935). Primarily for the purpose of comparing LMRL with intuitionistic linear logic and classical linear logic, we also present a variant of $\pi$-calculus for multiparty sessions that demonstrates a tight correspondence between process communication in this variant and multi-cut elimination in LMRL, thus extending some recent results on encoding session types as propositions in linear logic (Caires and Pfenning 2010; Wadler 2012a).

## 2. Linear Multirole Logic

While the first and foremost inspiration for multirole logic stems from studies on multiparty session types in distributed programming, it seems natural in retrospective to (also) introduce multirole logic by exploring (in terms of a notion referred to as role-based interpretation) the well-known duality between conjunction and disjunction in classical logic. For instance, in a two-sided presentation of the classical sequent calculus (LK), we have the following rules for conjunction and disjunction:

$$\frac{\underline{A} \vdash \underline{B}, A \quad \underline{A} \vdash \underline{B}, B}{\underline{A} \vdash \underline{B}, A \wedge B}\ \textbf{(conj-r)}$$

$$\frac{\underline{A}, A \vdash \underline{B}}{\underline{A}, A \wedge B \vdash \underline{B}}\ \textbf{(conj-l-1)} \qquad \frac{\underline{A}, B \vdash \underline{B}}{\underline{A}, A \wedge B \vdash \underline{B}}\ \textbf{(conj-l-2)}$$

$$\frac{\underline{A}, A \vdash \underline{B} \quad \underline{A}, B \vdash \underline{B}}{\underline{A}, A \vee B \vdash \underline{B}}\ \textbf{(disj-l)}$$

$$\frac{\underline{A} \vdash \underline{B}, A}{\underline{A} \vdash \underline{B}, A \vee B}\ \textbf{(disj-r-1)} \qquad \frac{\underline{A} \vdash \underline{B}, B}{\underline{A} \vdash \underline{B}, A \vee B}\ \textbf{(disj-r-2)}$$

where $\underline{A}$ and $\underline{B}$ range over sequents (that are essentially sequences of formulas). One possibility to explain this duality is to think of the availability of two roles 0 and 1 such that the left side of a sequent judgment (of the form $\underline{A} \vdash \underline{B}$) plays role 1 while the right side does role 0. In addition, there are two logical connectives $\wedge_0$ and $\wedge_1$; $\wedge_r$ is given a conjunction-like interpretation by the side playing role $r$ and disjunction-like interpretation by the other side playing role $1-r$, where $r$ ranges over 0 and 1. With this explanation, it seems entirely natural for us to introduce more roles into classical logic.

Given a natural number $N$, we use $\mathrm{R}_N$ for the set consisting of all of the natural numbers less than $N$, and $\mathrm{R}_\omega$ for the set of natural numbers. In addition, we use $\mathcal{R}$ for either $\mathrm{R}_N$ (for some $n \geq 2$) or $\mathrm{R}_\omega$, and may refer to each number in $\mathcal{R}$ as a role. Note that multirole logic is parameterized over a chosen underlying set $\mathcal{R}$ of roles, and we may use $\overline{\emptyset}$ to refer to this set $\mathcal{R}$. Given a subset $R$ of some $\mathcal{R}$, we use $\overline{R}$ for the complement of $R$ in $\mathcal{R}$ (assuming that this particular $\mathcal{R}$ can be readily inferred from the context). Also, we use $R_1 \uplus R_2$ for the union of two disjoint sets $R_1$ and $R_2$..

Intuitively speaking, a conjunctive multirole logic is one in which there is an underlying base set $\mathcal{R}$ of roles; for each $r \in \mathcal{R}$, there is a logical connective $\wedge_r$ such that $\wedge_r$ is given a conjunction-like interpretation by a side playing role $r$ and a disjunction-like interpretation otherwise. If we think of the universal quantifier $\forall$ as an infinite form of conjunction, then what is said about $\wedge$ can be readily applied to $\forall$ as well. In fact, additive, multiplicative, and exponential connectives in linear logic (Girard 1987) can all be treated in a similar manner. Evidently, a disjunctive multirole logic can be formulated dually (by giving $\wedge_r$ a disjunction-like interpretation if the side plays the role $r$ and a conjunction-like interpretation otherwise). While there is certainly a version of multirole logic based on classical logic, we solely focus on linear multirole logic (LMRL) in this paper, which is based on classical linear logic.

Given a formula $A$ and a set $R$ of roles (which is a subset of the underlying full set $\mathcal{R}$), we write $[A]_R$ for an i-formula, which is some sort of interpretation of $A$ based on $R$. For instance, the interpretation of $\wedge_r$ based on $R$ is conjunction-like if $r \in R$ holds, and it is disjunction-like otherwise. It is crucial to realize that interpretations should be based on sets of roles rather than just individual roles. In other words, one side is allowed to play multiple roles simultaneously.

A sequent $\Gamma$ in multirole logic is a sequence of i-formulas, and such a sequent is inherently many-sided as each $R$ appearing in $\Gamma$ represents just one side. Note that two identical i-formulas are allowed to appear in one sequent. We use $\emptyset$ for the empty sequence and $(\Gamma, [A]_R)$ for any sequence that can be formed by inserting $[A]_R$ into $\Gamma$ (at any position). The parentheses in $(\Gamma, [A]_R)$ may be dropped if there is no risk of confusion. We use $\vdash \Gamma$ for a judgment meaning that $\Gamma$ is derivable and may write $(\Gamma_1; \ldots; \Gamma_n) \Rightarrow \Gamma_c$ for an inference rule of the following form:

$$\frac{\vdash \Gamma_1 \quad \ldots \quad \vdash \Gamma_n}{\vdash \Gamma_c}$$

where $\vdash \Gamma_1, \ldots, \vdash \Gamma_n$ are the premisses of the rule and $\vdash \Gamma_c$ the conclusion.

As can be readily expected, the cut-rule (for two sequents) in (either conjunctive or disjunctive) LMRL is of the following form:

$$\frac{\Gamma_1, [A]_R \quad \Gamma_2, [A]_{\overline{R}}}{\Gamma_1, \Gamma_2}$$

The cut-rule can be interpreted as some sort of communication between two parties in distributed programming (Abramsky 1994a; Bellin and Scott 1994b; Caires and Pfenning 2010; Wadler 2012a). For communication between multiple parties, it is natural to seek a generalization of the cut-rule that involve more than two sequents. In conjunctive LMRL, the admissibility of the following cut-rule (*n-cut-conj*) can be established for each $n \geq 1$:

$$\frac{\overline{R}_1 \uplus \ldots \uplus \overline{R}_n = \overline{\emptyset} \quad \Gamma_1, [A]_{R_1} \quad \ldots \quad \Gamma_n, [A]_{R_n}}{\Gamma_1, \ldots, \Gamma_n}$$

In disjunctive LMRL, the admissibility of the following cut-rule (*n-cut-disj*) can be established for each $n \geq 1$:

$$\frac{R_1 \uplus \ldots \uplus R_n = \overline{\emptyset} \quad \Gamma_1, [A]_{R_1} \quad \ldots \quad \Gamma_n, [A]_{R_n}}{\Gamma_1, \ldots, \Gamma_n}$$

We will give explanation later on the case where $n = 1$. The case where $n = 2$ is special as both of the conditions $\overline{R}_1 \uplus \overline{R}_2 = \overline{\emptyset}$ $R_1 \uplus R_2 = \overline{\emptyset}$ are equivalent to $R_1$ and $R_2$ being complement to each other. Therefore, the rules *2-cut-conj* and *2-cut-disj* have the same form as the standard cut-rule (for two sequents). If $n$ is not 2, then the rules *n-cut-conj* and *n-cut-disj* impose different pre-conditions on the involved role sets $R_1, \ldots, R_n$. Also, please note that the pre-condition on $R_1, \ldots, R_n$ as is imposed by the rule *n-cut-conj* is identical to the requirement on $R_1, \ldots, R_n$ for linking $n$ matching channels of types $\mathbf{chan}(R_1, S), \ldots, \mathbf{chan}(R_1, S)$ stated in Section 1, which naturally prompts one to guess the existence of a profound relation between these two.

### 2.1 Syntax

We use $t$ for (first-order) terms in LMRL, which are standard. For each $r \in R$, there exist logical connectives $\otimes_r$, $\&_r$, $!_r$, and $\forall_r$. The formulas in LMRL are defined as follows:

$$\text{formulas} \quad A \quad ::= \quad a \mid A_1 \otimes_r A_2 \mid A_1 \&_r A_2 \mid !_r(A) \mid \forall_r(\lambda x.A)$$

where $a$ ranges over primitive ones. In CLL, $\otimes$ stands for the multiplicative conjunction, & the additive conjunction, ! the of-course modality operator, and $\forall$ the universal quantifier. An i-formula in LMRL is of the form $[A]_R$, and a sequent $\Gamma$ is a sequence of i-formulas. We may write $[?(A)]_R$ to mean $[!_r(A)]_R$ for some $r \notin R$, and $?(\Gamma)$ to mean that each i-formula in $\Gamma$ is of the form $[?(A)]_R$. Given a sequent $\Gamma$ and an i-formula $[A]_R$, we use $(\Gamma, [A]_R)$ for a sequent obtained from inserting $[A]_R$ into $\Gamma$ (at any position). Given two sequents $\Gamma_1$ and $\Gamma_2$, we use $(\Gamma_1, \Gamma_2)$ for a sequent obtained from merging $\Gamma_1$ with $\Gamma_2$ (in any kind of order). The parentheses in $(\Gamma, [A]_R)$ and $(\Gamma_1, \Gamma_2)$ may be dropped if there is no risk of confusion. Given an i-formula $[A]_R$, let us use $\{[A]_R\}$ for a sequent consisting of only $[A]_R$ if $A$ is not of the form $!_r(B)$ for $r \notin R$ or some repeated occurrences of $[A]_R$ otherwise (that is, if $A$ is of the form $!_r(B)$ for $r \notin R$).

### 2.2 LMRL$_\wedge$: Conjunctive LMRL

The inference rules for LMRL$_\wedge$ are given in Figure 2. Note that $\otimes_r$ is interpreted as $\otimes$ by a side playing the role $r$ and ⅋ (the dual of $\otimes$ in CLL) by a side not playing the role $r$; $\&_r$ is interpreted as & by a side playing the role $r$ and $\oplus$ (the dual of & in CLL) by a side not playing the role $r$; $!_r$ is interpreted as ! by a side playing the role $r$ and ? (the dual of ! in CLL) by a side not play the role $r$; $\forall_r$ is interpreted as $\forall$ (universal quantifier) by a side playing the role $r$ and $\exists$ (existential quantifier) by a side not playing the role $r$.

$$\frac{R_1 \uplus \ldots \uplus R_n = \overline{\emptyset}}{\vdash [a]_{R_1}, \ldots, [a]_{R_n}} \ (\mathbf{Id}_\wedge)$$

$$\frac{r \notin R \quad \vdash \Gamma, [A]_R, [B]_R}{\vdash \Gamma, [A \otimes_r B)]_R} \ (\otimes\text{-}\mathbf{neg})$$

$$\frac{r \in R \quad \vdash \Gamma_1, [A]_R \quad \vdash \Gamma_2, [B]_R}{\vdash \Gamma_1, \Gamma_2, [A \otimes_r B]_R} \ (\otimes\text{-}\mathbf{pos})$$

$$\frac{r \notin R \quad \vdash \Gamma, [A]_R}{\vdash \Gamma, [A \&_r B]_R} \ (\&\text{-}\mathbf{neg}\text{-}\mathbf{l})$$

$$\frac{r \notin R \quad \vdash \Gamma, [B]_R}{\vdash \Gamma, [A \&_r B]_R} \ (\&\text{-}\mathbf{neg}\text{-}\mathbf{r})$$

$$\frac{r \in R \quad \vdash \Gamma, [A]_R \quad \vdash \Gamma, [B]_R}{\vdash \Gamma, [A \&_r B]_R} \ (\&\text{-}\mathbf{pos})$$

$$\frac{r \in R \quad \vdash ?(\Gamma), [A]_R}{\vdash ?(\Gamma), [!_r(A)]_R} \ (!\text{-}\mathbf{pos})$$

$$\frac{r \notin R \quad \vdash \Gamma}{\vdash \Gamma, [!_r(A)]_R} \ (!\text{-}\mathbf{neg}\text{-}\mathbf{weaken})$$

$$\frac{r \notin R \quad \vdash \Gamma, [A]_R}{\vdash \Gamma, [!_r(A)]_R} \ (!\text{-}\mathbf{neg}\text{-}\mathbf{derelict})$$

$$\frac{r \notin R \quad \vdash \Gamma, [!_r(A)]_R, [!_r(A)]_R}{\vdash \Gamma, [!_r(A)]_R} \ (!\text{-}\mathbf{neg}\text{-}\mathbf{contract})$$

$$\frac{r \notin R \quad \vdash \Gamma, [A\{t/x\}]_R}{\vdash \Gamma, [\forall_r(\lambda x.A)]_R} \ (\forall\text{-}\mathbf{neg})$$

$$\frac{r \in R \quad x \notin \Gamma \quad \vdash \Gamma, [A]_R}{\vdash \Gamma, [\forall_r(\lambda x.A)]_R} \ (\forall\text{-}\mathbf{pos})$$

**Figure 2.** The inference rules for LMRL$_\wedge$

Note that there are one positive rule and one negative rule for each $\otimes_r$, and one positive rule and two negative rules for each $\&_r$, and one positive rule and one negative rule for each $\forall_r$. Let us take the rule **($\otimes$-neg)** as an example; the i-formula $[A \otimes_r B]_R$ is referred to as the major i-formula of the rule. Let us take the rule **($\otimes$-pos)** as another example; the i-formula $[A \otimes_r B]_R$ is referred to as the major i-formula of the rule. The major i-formulas for the other rules (excluding the rule **(Id$_\wedge$)**) should be clear as well. For the rule **(Id$_\wedge$)**, each $[a]_{R_i}$ is referred to as a major i-formula.

We use $|A|$ for the size of $A$, which is the number of connectives contained in $A$ We use $\mathcal{D}$ for a derivation tree and $ht(\mathcal{D})$ for the height of the tree. Also, we use $\mathcal{D} :: \Gamma$ for a derivation of $\Gamma$.

**Proposition 2.1** (Substitution)**.** *Given a sequent $\Gamma$, a variable $x$ and a term $t$, we use $\Gamma\{x/t\}$ for the sequent obtained from replacing each i-formula $[A]_R$ in $\Gamma$ with $[A\{x/t\}]_R$. Assume $\mathcal{D}_1 :: \Gamma$. Then we can construct a derivation $\mathcal{D}_2$ of the sequent $\Gamma\{x/t\}$.*

*Proof.* By structural induction on $\mathcal{D}_1$. □

We may use $\mathcal{D}_1\{x/t\}$ for the $\mathcal{D}_2$ constructed in Proposition 2.1.

**Lemma 2.2.** *The following rule is admissible in LMRL$_\wedge$:*

$$(\Gamma, [A]_\emptyset) \Rightarrow \Gamma$$

*Proof.* By structural induction on the derivation of $\mathcal{D} :: (\Gamma, [A]_\emptyset)$. □

Note that Lemma 2.2 simply states the admissibility of the cut-rule *n-cut-conj* for $n = 1$. So it actually makes sense to have a cut involving only one sequent!

**Lemma 2.3** ($\eta$-expansion)**.** *The following rule is admissible in* $LMRL_\wedge$*:*

$$() \Rightarrow [A]_{R_1}, \ldots, [A]_{R_n}$$

*where* $R_1 \uplus \ldots \uplus R_n = \overline{\emptyset}$.

*Proof.* By structural induction on $A$. □

The next lemma is the most crucial one in this paper. While its proof may seem rather involved, it should be readily accessible for someone familiar with a standard cut-elimination proof.

**Lemma 2.4** (2-cut with spill)**.** *Assume that* $\overline{R}_1$ *and* $\overline{R}_2$ *are disjoint. Then the following rule is admissible in* $LMRL_\wedge$*:*

$$(\Gamma_1, [A]_{R_1}; \Gamma_2, [A]_{R_2}) \Rightarrow \Gamma_1, \Gamma_2, [A]_{R_1 \cap R_2}$$

*Proof.* Due to the explicit presence of the three structural rules **(!-neg-weaken)** **(!-neg-derelict)**, and **(!-neg-contract)** in $LMRL_\wedge$, we need to prove a strengthened version of Lemma 2.4 stating that the following rule is admissible in $LMRL_\wedge$:

$$(\Gamma_1, \{[A]_{R_1}\}; \Gamma_2, \{[A]_{R_2}\}) \Rightarrow \Gamma_1, \Gamma_2, [A]_{R_1 \cap R_2}$$

Note that the proof strategy we use is essentially adopted from the one in a proof of cut-elimination for classical linear logic (CLL) (Troelstra 1992). Assume $\mathcal{D}_1 :: (\Gamma_1, \{[A]_{R_1}\})$ and $\mathcal{D}_2 :: (\Gamma_2, \{[A]_{R_2}\})$. We proceed by induction on $|A|$ (the size of $A$) and $ht(\mathcal{D}_1) + ht(\mathcal{D}_2)$, lexicographically ordered. For brevity, we are to focus only on the most interesting case where there is one occurrence of $[A]_{R_i}$ in $\{[A]_{R_i}\}$ that is the major formula of the last rule applied in $\mathcal{D}_i$, where $i$ ranges over 1 and 2. For this case, we have several subcases covering all the possible forms that $A$ may take.

**Assume that** $A$ **is primitive**. Then it is a simple routine to verify that the sequent $\vdash (\Gamma_1, \Gamma_2, [A]_{R_1 \cap R_2})$ follows from an application of the rule $(\mathbf{Id}_\wedge)$.

**Assume that** $A$ **is of the form** $A_1 \otimes_r A_2$. We have three possibilities: $r \in R_1$ and $r \notin R_2$, or $r \notin R_1$ and $r \in R_2$, or $r \in R_1$ and $r \in R_2$.

- Assume $r \in R_1$ and $r \notin R_2$. Then $\mathcal{D}_1$ is of the following form:

$$\frac{\mathcal{D}_{11} :: (\Gamma_{11}, [A_1]_{R_1}) \quad \mathcal{D}_{12} :: (\Gamma_{12}, [A_2]_{R_1})}{\vdash \Gamma_1, [A]_{R_1}} \ (\otimes\textbf{-pos})$$

  and $\mathcal{D}_2$ is of the following form:

$$\frac{\mathcal{D}_{21} :: (\Gamma_2, [A_1]_{R_2}, [A_2]_{R_2})}{\vdash \Gamma_2, [A]_{R_2}} \ (\otimes\textbf{-neg})$$

  By the induction hypothesis on $\mathcal{D}_{11}$ and $\mathcal{D}_{21}$, we have a derivation:

$$\mathcal{D}'_{11} :: (\Gamma_{11}, \Gamma_2, [A_1]_{R_1 \cap R_2}, [A_2]_{R_2})$$

  By the induction hypothesis on $\mathcal{D}_{12}$ and $\mathcal{D}'_{11}$, we have a derivation:

$$\mathcal{D}'_{12} :: (\Gamma, [A_1]_{R_1 \cap R_2}, [A_2]_{R_1 \cap R_2})$$

  By applying the rule **(⊗-neg)** to $\mathcal{D}'_{12}$, we have a derivation of the sequent $(\Gamma, [A]_{R_1 \cap R_2})$.
- Assume $r \notin R_1$ and $r \in R_2$. Then this case is analogous to the previous one.
- Assume $r \in R_1$ and $r \in R_2$. Then $\mathcal{D}_k$ is of the following form for each of the cases $k = 1$ and $k = 2$:

$$\frac{\mathcal{D}_{k1} :: (\Gamma_{k1}, [A_1]_{R_k}) \quad \mathcal{D}_{k2} :: (\Gamma_{k2}, [A_2]_{R_k})}{\vdash \Gamma_k, [A]_{R_k}} \ (\otimes\textbf{-pos})$$

  By the induction hypothesis on $\mathcal{D}_{11}$ and $\mathcal{D}_{21}$, we obtain a derivation:

$$\mathcal{D}'_1 :: (\Gamma_{11}, \Gamma_{21}, [A_1]_{R_1 \cap R_2})$$

  By the induction hypothesis on $\mathcal{D}_{12}$ and $\mathcal{D}_{22}$, we obtain a derivation:

$$\mathcal{D}'_2 :: (\Gamma_{12}, \Gamma_{22}, [A_2]_{R_1 \cap R_2})$$

  By applying the rule **(⊗-pos)** to $\mathcal{D}'_1$ and $\mathcal{D}'_2$, we obtain a derivation of the sequent $(\Gamma, [A]_{R_1 \cap R_2})$.

**Assume that** $A$ **is of the form** $A_1 \,\&_r\, A_2$. We have three possibilities: $r \in R_1$ and $r \notin R_2$, or $r \notin R_1$ and $r \in R_2$, or $r \in R_1$ and $r \in R_2$.

- Assume $r \in R_1$ and $r \notin R_2$. Then $\mathcal{D}_1$ is of the following form:

$$\frac{\mathcal{D}_{11} :: (\Gamma_1, [A_1]_{R_1}) \quad \mathcal{D}_{12} :: (\Gamma_1, [A_2]_{R_1})}{\vdash \Gamma_1, [A]_{R_1}} \ (\&\textbf{-pos})$$

  and $\mathcal{D}_2$ is of the following form for $k$ being either 1 or 2:

$$\frac{\mathcal{D}_{2k} :: (\Gamma_2, [A_k]_{R_2})}{\vdash \Gamma_2, [A]_{R_2}}$$

  where the last applied rule in $\mathcal{D}_2$ is **(&-neg-l)** or **(&-neg-r)**. By induction hypothesis on $\mathcal{D}_{1k}$ and $\mathcal{D}_{2k}$, we obtain a derivation:

$$\mathcal{D}'_k :: (\Gamma_1, \Gamma_2, [A_k]_{R_1 \cap R_2})$$

  By applying to $\mathcal{D}'_k$ either **(&-neg-l)** or **(&-neg-r)**, we obtain a derivation of the sequent $(\Gamma_1, \Gamma_2, [A]_{R_1 \cap R_2})$.
- Assume $r \notin R_1$ and $r \in R_2$. Then this case is analogous to the previous one.
- Assume $r \in R_1$ and $r \in R_2$. Then $\mathcal{D}_k$ is of the following form for each of the cases $k = 1$ and $k = 2$:

$$\frac{\mathcal{D}_{k1} :: (\Gamma_{k1}, [A_1]_{R_k}) \quad \mathcal{D}_{k2} :: (\Gamma_{k2}, [A_2]_{R_k})}{\vdash \Gamma_k, [A]_{R_k}} \ (\&\textbf{-pos})$$

  By the induction hypothesis on $\mathcal{D}_{11}$ and $\mathcal{D}_{21}$, we obtain a derivation:

$$\mathcal{D}'_1 :: (\Gamma_{11}, \Gamma_{21}, [A_1]_{R_1 \cap R_2})$$

  By the induction hypothesis on $\mathcal{D}_{12}$ and $\mathcal{D}_{22}$, we obtain a derivation:

$$\mathcal{D}'_2 :: (\Gamma_{12}, \Gamma_{22}, [A_2]_{R_1 \cap R_2})$$

  By applying the rule **(&-pos)** to $\mathcal{D}'_1$ and $\mathcal{D}'_2$, we obtain a derivation of the sequent $(\Gamma_1, \Gamma_2, [A]_{R_1 \cap R_2})$.

**Assume that** $A$ **is of the form** $!_r(B)$. This is the most involved subcase. We have three possibilities: $r \in R_1$ and $r \notin R_2$, or $r \notin R_1$ and $r \in R_2$, or $r \in R_1$ and $r \in R_2$.

- Assume $r \in R_1$ and $r \notin R_2$. Then $\mathcal{D}_1$ is of the following form:

$$\frac{\mathcal{D}_{11} :: ?(\Gamma_1), [B]_{R_1}}{\vdash ?(\Gamma_1), [A]_{R_1}} \ (!\textbf{-pos})$$

  There are the following three possibilities for $\mathcal{D}_2$:
  - $\mathcal{D}_2$ is of the following form:

$$\frac{\mathcal{D}_{21} :: \Gamma_2, \{[A]_{R_2}\}}{\vdash \Gamma_2, \{[A]_{R_2}\}} \ (!\textbf{-neg-weaken})$$

    We simply obtain a derivation of $(\Gamma_1, \Gamma_2, [A]_{R_1 \cap R_2})$ by the induction hypothesis on $\mathcal{D}_1$ and $\mathcal{D}_{21}$.
  - $\mathcal{D}_2$ is of the following form:

$$\frac{\mathcal{D}_{21} :: \Gamma_2, \{[A]_{R_2}\}, [B]_R}{\vdash \Gamma_2, \{[A]_{R_2}\}} \ (!\textbf{-neg-derelict})$$

By the induction hypothesis on $\mathcal{D}_1$ and $\mathcal{D}_{21}$, we obtain a derivation:

$$\mathcal{D}_{121} :: (\Gamma_1, \Gamma_2, [A]_{R_1 \cap R_2}, [B]_R)$$

By the induction hypothesis on $\mathcal{D}_{11}$ and $\mathcal{D}_{121}$, we obtain a derivation:

$$\mathcal{D}'_{121} :: (\Gamma_1, \Gamma_1, \Gamma_2, [A]_{R_1 \cap R_2})$$

By applying the rule **(!-neg-contract)** to $\mathcal{D}'_{121}$ repeatedly, we obtain a derivation of $(\Gamma_1, \Gamma_2, [A]_{R_1 \cap R_2})$.

- $\mathcal{D}_2$ is of the following form:

$$\frac{\mathcal{D}_{21} :: \Gamma_2, \{[A]_{R_2}\}, [A]_{R_2}}{\vdash \Gamma_2, \{[A]_{R_2}\}} \ \textbf{(!-neg-contract)}$$

We simply obtain a derivation of $(\Gamma_1, \Gamma_2, [A]_{R_1 \cap R_2})$ by the induction hypothesis on $\mathcal{D}_1$ and $\mathcal{D}_{21}$.

- Assume $r \notin R_1$ and $r \in R_2$. This subcase is completely analogous to the previous one.
- Assume $r \in R_1$ and $r \in R_2$. Then $\mathcal{D}_k$ is of the following form for each of the cases $k = 1$ and $k = 2$:

$$\frac{\mathcal{D}_{k1} :: ?(\Gamma_k), [B]_{R_k}}{\vdash ?(\Gamma_k), [A]_{R_k}} \ \textbf{(!-pos)}$$

We obtain $\mathcal{D}'_{12} :: (?(\Gamma_1), ?(\Gamma_2), [B]_{R_1 \cap R_2})$ by the induction hypothesis on $\mathcal{D}_{11}$ and $\mathcal{D}_{21}$. We then obtain a derivation of $(?(\Gamma_1), ?(\Gamma_2), [A]_{R_1 \cap R_2})$ by applying the rule **(!-pos)** to $\mathcal{D}'_{12}$.

**Assume** $A$ **is of the form** $\forall_r(\lambda x.B)$. We have three possibilities: $r \in R_1$ and $r \notin R_2$, or $r \in R_2$ and $r \notin R_1$, or $r \in R_1$ and $r \in R_2$.

- Assume $r \in R_1$ and $r \notin R_2$. Then $\mathcal{D}_1$ is of the following form:

$$\frac{\mathcal{D}_{11} :: (\Gamma_1, [B]_{R_1})}{\vdash \Gamma, [A]_{R_1}} \ \textbf{(}\forall\textbf{-pos)}$$

where $x$ does not have any free occurrences in $\Gamma_1$, and $\mathcal{D}_2$ is of the following form:

$$\frac{\mathcal{D}_{21} :: (\Gamma_2, [B\{x/t\}]_{R_2})}{\vdash \Gamma, [A]_{R_2}} \ \textbf{(}\forall\textbf{-neg)}$$

Let $\mathcal{D}'_{11}$ be $\mathcal{D}_{11}\{x/t\}$, which is a derivation of $(\Gamma_1, [B\{x/t\}]_{R_1})$. By the induction hypothesis on $\mathcal{D}'_{11}$ and $\mathcal{D}_{21}$, we have a derivation:

$$\mathcal{D}_{121} :: (\Gamma_1, \Gamma_2, [B\{x/t\}]_{R_1 \cap R_2})$$

By applying the rule **(∀-neg)** to $\mathcal{D}_{121}$, we have a derivation of $(\Gamma_1, \Gamma_2, [A]_{R_1 \cap R_2})$.

- Assume $r \notin R_1$ and $r \in R_2$. Then this case is analogous to the previous one.
- Assume $r \in R_1$ and $r \in R_2$. Then $\mathcal{D}_k$ is of the following form for each of the cases $k = 1$ and $k = 2$:

$$\frac{\mathcal{D}_{k1} :: (\Gamma_k, [A]_{R_k}, [B]_{R_k})}{\vdash \Gamma_k, [A]_{R_k}} \ \textbf{(}\forall\textbf{-pos)}$$

where $x$ does not have free occurrences in $\Gamma_k$. By the induction hypothesis on $\mathcal{D}_{11}$ and $\mathcal{D}_{21}$, we have a derivation:

$$\mathcal{D}'_{12} :: (\Gamma_1, \Gamma_2, [B]_{R_1 \cap R_2})$$

By applying the rule **(∀-pos)** to $\mathcal{D}'_{12}$, we obtain a derivation of $(\Gamma_1, \Gamma_2, [A]_{R_1 \cap R_2})$.

All of the cases are covered where the cut-formula is the major formula of both $\mathcal{D}_1$ and $\mathcal{D}_2$. For brevity, we omit the cases where the cut-formula is not the major formula of either $\mathcal{D}_1$ or $\mathcal{D}_2$, which can be trivially handled (Troelstra 1992). □

**Lemma 2.5** (2-cut)**.** *The following rule is admissible in LMRL*$_\wedge$*:*

$$(\Gamma_1, [A]_R; \Gamma_2, [A]_{\overline{R}}) \Rightarrow \Gamma_1, \Gamma_2$$

Lemma 2.5 is just a special case of Lemma 2.6 where $n$ is 2.

**Lemma 2.6** ($n$-cut)**.** *Assume that* $R_1, R_2, \ldots, R_n$ *are subsets of* $\mathcal{R}$ *for some* $n \geq 1$ *such that:*

$$\overline{R}_1 \uplus \overline{R}_2 \uplus \ldots \uplus \overline{R}_n = \overline{\emptyset}$$

*Then the following rule is admissible in LMRL*$_\wedge$*:*

$$(\Gamma_1, [A]_{R_1}; \Gamma_2, [A]_{R_2}; \ldots; \Gamma_n, [A]_{R_n}) \Rightarrow \Gamma_1, \Gamma_2, \ldots, \Gamma_n$$

*Proof.* Assume $\mathcal{D}_i :: (\Gamma_1, [A]_{R_i})$ for $1 \leq i \leq n$. The proof proceeds by induction on $n$. The base case (where $n = 1$) is simply covered by Lemma 2.2. For $n \geq 2$, we can apply Lemma 2.4 to $\mathcal{D}_1$ and $\mathcal{D}_2$ to obtain a derivation $\mathcal{D}_{12} :: (\Gamma_1, \Gamma_2, [A]_{R_{12}})$ where $R_{12} = R_1 \cap R_2$. Clearly, $\overline{R}_{12} = \overline{R}_1 \uplus \overline{R}_2$ holds, and we can invoke induction hypothesis on $\mathcal{D}_{12}$ and the remaining derivations $\mathcal{D}_i$ for $3 \leq i \leq n$ to obtain a derivation of $(\Gamma_1, \ldots, \Gamma_n)$. □

**Lemma 2.7** (Splitting)**.** *The following rule is admissible in LMRL*$_\wedge$*:*

$$(\Gamma, [A]_{R_1 \uplus R_2}) \Rightarrow \Gamma, [A]_{R_1}, [A]_{R_2}$$

*Proof.* Assume $\mathcal{D}_1 :: (\Gamma, [A]_{R_1 \uplus R_2})$. By Lemma 2.3, we have a derivation:

$$\mathcal{D}_2 :: ([A]_{\overline{R}_1 \cap \overline{R}_2}, [A]_{R_1}, [A]_{R_2})$$

By applying Lemma 2.5 to $\mathcal{D}_1$ and $\mathcal{D}_2$, we obtain a derivation of $(\Gamma, [A]_{R_1}, [A]_{R_2})$. □

### 2.3 LMRL$_\vee$ as the Dual of LMRL$_\wedge$: Disjunctive LMRL

For a bit of completeness, we introduce disjunctive LMRL (LMRL$_\vee$), which is the exact dual of LMRL$_\wedge$. The inference rules for LMRL$_\vee$ are listed in Figure 3, which can simply be obtained by replacing each set (of roles) in Figure 2 with its complement. The various lemmas established for LMRL$_\wedge$ are given their counterparts in LMRL$_\vee$ as follows:

**Lemma 2.8.** *The following rule is admissible in LMRL*$_\vee$*:*

$$(\Gamma, [A]_{\overline{\emptyset}}) \Rightarrow \Gamma$$

Note that Lemma 2.8 simply states the admissibility of the cut-rule *n-cut-disj* for $n = 1$.

**Lemma 2.9** ($\eta$-expansion)**.** *The following rule is admissible in LMRL*$_\vee$*:*

$$() \Rightarrow [A]_{R_1}, \ldots, [A]_{R_n}$$

*where* $\overline{R}_1 \uplus \ldots \uplus \overline{R}_n = \overline{\emptyset}$*.*

**Lemma 2.10** (2-cut)**.** *The following rule is admissible in LMRL*$_\vee$*:*

$$(\Gamma_1, [A]_R; \Gamma_2, [A]_{\overline{R}}) \Rightarrow \Gamma_1, \Gamma_2$$

**Lemma 2.11** (2-cut with spill)**.** *Assume that* $R_1$ *and* $R_2$ *are disjoint. Then the following rule is admissible in LMRL*$_\vee$*:*

$$(\Gamma_1, [A]_{R_1}; \Gamma_2, [A]_{R_2}) \Rightarrow \Gamma_1, \Gamma_2, [A]_{R_1 \uplus R_2}$$

**Lemma 2.12** (Multi-cut)**.** *Assume that* $R_1, R_2, \ldots, R_n$ *are subsets of* $\mathcal{R}$ *for some* $n \geq 2$ *such that:*

$$R_1 \uplus R_2 \uplus \ldots \uplus R_n = \overline{\emptyset}$$

*Then the following rule is admissible in LMRL*$_\vee$*:*

$$(\Gamma_1, [A]_{R_1}; \Gamma_2, [A]_{R_2}; \ldots; \Gamma_n, [A]_{R_n}) \Rightarrow \Gamma_1, \Gamma_2, \ldots, \Gamma_n$$

$$\frac{\overline{R}_1 \uplus \dots \uplus \overline{R}_n = \overline{\emptyset}}{\vdash [a]_{R_1}, \dots, [a]_{R_n}}\ (\mathbf{Id}_\vee)$$

$$\frac{r \in R \quad \vdash \Gamma, [A]_R, [B]_R}{\vdash \Gamma, [A \otimes_r B)]_R}\ (⅋\textbf{-pos})$$

$$\frac{r \notin R \quad \vdash \Gamma_1, [A]_R \quad \vdash \Gamma_2, [B]_R}{\vdash \Gamma_1, \Gamma_2, [A \otimes_r B]_R}\ (⅋\textbf{-neg})$$

$$\frac{r \in R \quad \vdash \Gamma, [A]_R}{\vdash \Gamma, [A \,\&_r\, B]_R}\ (\oplus\textbf{-pos-l})$$

$$\frac{r \in R \quad \vdash \Gamma, [B]_R}{\vdash \Gamma, [A \,\&_r\, B]_R}\ (\oplus\textbf{-pos-r})$$

$$\frac{r \notin R \quad \vdash \Gamma, [A]_R \quad \vdash \Gamma, [B]_R}{\vdash \Gamma, [A \,\&_r\, B]_R}\ (\oplus\textbf{-neg})$$

$$\frac{r \notin R \quad \vdash ?(\Gamma), [A]_R}{\vdash ?(\Gamma), [!_r(A)]_R}\ (?\textbf{-neg})$$

$$\frac{r \in R \quad \vdash \Gamma}{\vdash \Gamma, [!_r(A)]_R}\ (?\textbf{-pos-weaken})$$

$$\frac{r \in R \quad \vdash \Gamma, [A]_R}{\vdash \Gamma, [!_r(A)]_R}\ (?\textbf{-pos-derelict})$$

$$\frac{r \in R \quad \vdash \Gamma, [!_r(A)]_R, [!_r(A)]_R}{\vdash \Gamma, [!_r(A)]_R}\ (?\textbf{-pos-contract})$$

$$\frac{r \in R \quad \vdash \Gamma, [A\{t/x\}]_R}{\vdash \Gamma, [\forall_r(\lambda x.A)]_R}\ (\exists\textbf{-pos})$$

$$\frac{r \notin R \quad x \notin \Gamma \quad \vdash \Gamma, [A]_R}{\vdash \Gamma, [\forall_r(\lambda x.A)]_R}\ (\exists\textbf{-neg})$$

**Figure 3.** The inference rules for $\mathrm{LMRL}_\vee$

**Lemma 2.13** (Co-Splitting)**.** *The following rule is admissible in* $\mathit{LMRL}_\vee$*:*

$$(\Gamma, [A]_{R_1 \cap R_2}) \Rightarrow \Gamma, [A]_{R_1}, [A]_{R_2}$$

*where* $\overline{R}_1$ *and* $\overline{R}_2$ *are disjoint.*

The proofs for these lemmas (2.8, 2.10, 2.11, 2.12, and 2.13) are omitted as they are completely analogous to those for the corresponding lemmas in $\mathrm{LMRL}_\wedge$.

### 2.4 About LMRL and Negation

Given an i-formula $[A]_R$, we may think of the i-formula $[A]_{\overline{R}}$ as the negation of $[A]_R$. For the moment, let us use $\neg A$ for the negation of $A$ and introduce the following rule for handling negation:

$$\frac{\Gamma, [A]_R}{\Gamma, [\neg A]_{\overline{R}}}\ (\mathbf{not}_\wedge)$$

It is clear that $\mathrm{LMRL}_\wedge$ extended with the rule $(\mathbf{not}_\wedge)$ corresponds precisely to CLL if the underlying set $\overline{\emptyset}$ of roles equals $\{0, 1\}$. The logical connective $\wedge_0$ and $\wedge_1$ correspond to $\wedge$ and $\vee$, respectively, and the quantifiers $\forall_0$ and $\forall_1$ correspond to $\forall$ and $\exists$, respectively. Each sequent $\Gamma$ in $\mathrm{LMRL}_\wedge$ can be translated into a sequent $\underline{A}_1 \vdash \underline{A}_0$ as follows such that $\Gamma$ is derivable in $\mathrm{LMRL}_\wedge$ if and only if $\underline{A}_1 \vdash \underline{A}_0$ is derivable in CLL:

- If $[A]_{\overline{\emptyset}}$ is in $\Gamma$, then 1 goes into $\underline{A}_0$;
- If $[A]_{\emptyset}$ is in $\Gamma$, then 1 goes into $\underline{A}_1$;
- If $[A]_{\{0\}}$ is in $\Gamma$, then $A$ goes into $\underline{A}_0$;
- If $[A]_{\{1\}}$ is in $\Gamma$, then $A$ goes into $\underline{A}_1$.

| | | | |
|---|---|---|---|
| $P$ | $::=$ | $\nu x : A.(\vec{P})$ | linking |
| | | $x_R(y)_r.(P_1 \mid P_2)$ | name input ($r \notin R$) |
| | | $x_R[y]_r.P$ | name output ($r \in R$) |
| | | $x_R(_)_r.0$ | empty input ($r \notin R$) |
| | | $x_R[_]_r.P$ | empty output ($r \in R$) |
| | | $x_R(case)_r.(P_1, P_2)$ | case offer ($r \notin R$) |
| | | $x_R[inl]_r.P$ | left choice ($r \in R$) |
| | | $x_R[inr]^r.P$ | right choice ($r \in R$) |
| | | $!x_R(y)_r$ | server accept ($r \notin R$) |
| | | $?x_R[y]_r$ | client request ($r \in R$) |

**Figure 4.** The syntax for processes in $\pi$LMRL

Note that 1 is the unit of $\otimes$ in CLL. It is also clear that the presence of the rule $(\mathbf{not}_\wedge)$ invalidates the admissibility of $n$-cut for any $n \neq 2$. Let us take a look at the case where $n = 3$. Applying a 3-cut to $(\Gamma, [\neg A]_{R_1})$, $(\Gamma, [\neg A]_{R_2})$, and $(\Gamma, [\neg A]_{R_3})$ requires that the condition $\overline{R}_1 \uplus \overline{R}_2 \uplus \overline{R}_3 = \overline{\emptyset}$ be met; this 3-cut is supposed to be reduced to another 3-cut on $(\Gamma, [A]_{\overline{R}_1})$, $(\Gamma, [A]_{\overline{R}_2})$, and $(\Gamma, [A]_{\overline{R}_3})$; this new 3-cut is unfortunately not valid since it requires that the condition $R_1 \uplus R_2 \uplus R_3 = \overline{\emptyset}$ be met (which contradicts the previous condition $\overline{R}_1 \uplus \overline{R}_2 \uplus \overline{R}_3 = \overline{\emptyset}$).

We see LMRL as a form of negation-less logic: The negation of each i-formula $[A]_R$ is simply the i-formula $[A]_{\overline{R}}$ but there is no negation for the formula $A$ *per se*. In other words, the notion of negation cannot be internalized within LMRL if 3-cut is to be preserved.

## 3. LMRL as a Process Calculus

As the inspiration for LMRL stems from studies on multiparty session types, it only seems fit if we present a typed variant ($\pi$LMRL) of $\pi$-calculus (Milner et al. 1992c) in which the types are directly based on the formulas in LMRL. We are to closely follow some recent work by Caires and Pfenning (2010) and Wadler (2012) in our presentation of $\pi$LMRL. In particular, we shall mostly adopt the notational convention used by the latter and thus refer the reader to the original paper for detailed explanation.

We see the encoding of cut-elimination of LMRL in $\pi$LMRL mostly as a routine exercise. However, it should be noted that there exists a fundamental difference between $\pi$LMRL and $\pi$-calculus: The point-to-point communication in the latter is replaced with a form of broadcasting in the former. We are to mention at the end a simple extension of $\pi$LMRL that can support point-to-point communication directly.

### 3.1 Session Types

The session types (or types for short) are just formulas in LMRL except for adding $1_r$ (the unit for $\otimes_r$) for each role $r$ and dropping primitive formulas as well as quantified formulas:

$$\text{Session Types} \quad A \quad ::= \quad 1_r \mid A_1 \otimes_r A_2 \mid A_1 \,\&_r\, A_2 \mid !_r(A)$$

The meaning of various forms of session types is to be given later.

### 3.2 Process Terms

The syntax for the terms representing processes in $\pi$LMRL are given in Figure 3.1. We use $x$ and $y$ for names (of channels). Given $x$ and $R$, the term $x_R$ refers to an endpoint of the name $x$ such that the endpoint is supposed to be held by a party playing the roles contained in $R$. We use $P$ for a process and $\vec{P}$ for a sequence of the form $(P_1 \mid \dots \mid P_n)$ where $n \geq 1$. In $\nu x : A.(\vec{P})$, the name $x$ is bound in $\vec{P}$; in $x_R(y)_r.(P_1 \mid P_2)$, the name $y$ is bound in $P_1$ (but not in $P_2$); in $x_R[y]_r$, $!x_R(y)_r$, and $?x_R[y]_r$, the name $y$ is bound in $P$. Given a process $P$, we write $fn(P)$ for the set of free names in $P$. Note that the presence of annotations like $A$, $R$, and $r$ in process terms is solely for supporting a form of Church typing. Such annotations may be omitted if there is no risk of confusion. In particular, they are not needed in the formulation of reduction semantics for $\pi$LMRL.

Given $P_1, \ldots, P_n$ for some $n \geq 2$, $\nu x.(P_1 | \ldots | P_n)$, which is often referred to as a cut, means to link $n$ endpoints $x_{R_i}$ for some name $x$ and role sets $R_1, \ldots, R_n$ such that $\overline{R_1} \uplus \ldots \uplus \overline{R_n} = \overline{\emptyset}$ and each $x_{R_i}$ is contained in $P_i$ for $1 \leq i \leq n$.

The meaning of various headers in Figure 3.1 is to be made clear when the reduction semantcs for $\pi$LMRL is formulated. For the moment, we present a bit of intuition on them as follows.

A header term $x_R(y)_r$ means that $r \in R$ holds and an name is to be received on the endpoint $x_R$. Normally, substitution of the received name for $y$ would take place explicitly. As $y$ is a bound variable, another option, which we take here, is to rename $y$ implicitly to match the received name. Dually, a header term $x_R[y]_r$ means that $r \notin R$ holds and a fresh name is to be chosen and then sent onto the endpoint $x_R$.

A header term $x_R[_]_r$ means a send action is to take place on $x_R$ but no name is actually sent. Dually, a header term $x_R(_)_r$ means a recieve action is to take place on $x_R$ but no name is actually received.

A header term $x_R[inl]_r$ ($x_R[inr]_r$) means that a left (right) choice is to be sent onto the endpoint $x_R$. A header term $x_R(case)_r$ means that either a left or right choice is to be received on the endpoint $x_R$ and the received choice determines which of the two processes following the header should be chosen.

A header term $!x_R[y]_r$ essentially means that it can be repeatly used as $x_R[y]_r$, and a header term $?x_R(y)_r$ is like $x_R(y)_r$ but it is supposed to match $!x_R[y]_r$.

### 3.3 Statics of $\pi$LMRL

Given a type and a role set $R$, we can form an i-type $[A]_R$. If one thinks that $A$ is a global type, then $[A]_R$ is a local type for the endpoint of a full channel that is supposed to be held by a party playing the roles in $R$. The meaning for various forms of session types is to become clear when the typing rules in $\pi$LMRL are presented.

Let us write $[A \otimes B]_R$ ($[A \parr B]_R$) to mean $[A \otimes_r B]_R$ for some $r \in R$ ($r \notin R$). In the literature, $A \otimes B$ is often intuitively interpreted as *output A then behave as B* (Caires and Pfenning 2010; Wadler 2012a), but we are to see that it cannot be the case in $\pi$LMRL; it must be interpreted as *input A then behave as B* as is stipulated by the proof of cut-elimination. Dually, $A \parr B$ must be interpreted as *output A then behave as B* in $\pi$LMRL. Let us write $[A \& B]_R$ ($[A \oplus B]_R$) to mean $[A \&_r B]_R$ for some $r \in R$ ($r \notin R$); $A \& B$ intuitively means offering choice of $A$ or $B$ and $A \oplus B$ means selecting from $A$ or $B$. Let us write $[!A]_R$ ($[?A]_R$) to mean $[!_r A]_R$ for some $r \in A$ ($r \notin A$); $!A$ means the ability to repeatedly spawn processes of the type $A$ while $?A$ means to request such a process to be spawned.

In $\pi$LMRL, we use $\Gamma$ for an environment associating distinct names with i-types. Let $\Gamma$ be the following environment:

$$x_1 : [A_1]_{R_1}, \ldots, x_n : [A_n]_{R_n}$$

Then the domain $\mathbf{dom}(\Gamma)$ of $\Gamma$ equals $\{x_1, \ldots, x_n\}$. Given another environment $\Gamma'$, we write $\Gamma, \Gamma'$ for the union of $\Gamma$ and $\Gamma_1$ whenever $\mathbf{dom}(\Gamma) \cap \mathbf{dom}(\Gamma') = \emptyset$. We refer to $x : [A]_R$ as an association in an environment.

A typing judgment in $\pi$LMRL is of the form

$$P \vdash x_1 : [A_1]_{R_1}, \ldots, x_n : [A_n]_{R_n}$$

meaning that process $P$ communicates along each endpoint $(x_i)_{R_i}$ in a full channel specified by the session type $A_k$ for $k = 1, \ldots, n$. Erasing $P$ and the names $x_k$ from the judgment yields a judgment in LMRL (extended with $1_r$ (as the unit for $\otimes_r$) for each role $r$).

$$\frac{\overline{R_1} \uplus \ldots \uplus \overline{R_n} = \overline{\emptyset} \quad P_i \vdash \Gamma_i, x : [A]_{R_i} \text{ for } 1 \leq i \leq n}{\nu x.(P_1 | \ldots | P_n) \vdash \Gamma_1, \ldots, \Gamma_n} \ (\textbf{n-cut})$$

$$\frac{r \notin R \quad P \vdash \Gamma, y : [A_1]_R, x : [A_2]_R}{x[y]_r.P \vdash \Gamma, x : [A_1 \otimes_r A_2]_R} \ (\otimes\textbf{-neg})$$

$$\frac{r \in R \quad P_1 \vdash \Gamma_1, y : [A_1]_R \quad P_2 \vdash \Gamma_2, x : [A_2]_R}{x(y)_r.(P_1, P_2) \vdash \Gamma_1, \Gamma_2, x : [A_1 \otimes_r A_2]_R} \ (\otimes\textbf{-pos})$$

$$\frac{r \in R \quad P \vdash \Gamma}{x_R[_]_r.P \vdash \Gamma, x : [1_r]_R} \ (\textbf{1-neg}) \qquad \frac{r \in R}{x_R(_)_r.0 \vdash x : [1_r]_R} \ (\textbf{1-pos})$$

$$\frac{r \notin R \quad P \vdash \Gamma, x : [A_1]_R}{x_R[inl]_r.P \vdash \Gamma, x : [A_1 \&_r A_2]_R} \ (\&\textbf{-neg-l})$$

$$\frac{r \notin R \quad P \vdash \Gamma, x : [A_2]_R}{x_R[inr]_r.P \vdash \Gamma, x : [A_1 \&_r A_2]_R} \ (\&\textbf{-neg-r})$$

$$\frac{r \in R \quad P_1 \vdash \Gamma, x : [A_1]_R \quad P_2 \vdash \Gamma, x : [A_2]_R}{x_R(case)_r.(P_1, P_2) \vdash \Gamma, x : [A_1 \&_r A_2]_R} \ (\&\textbf{-pos})$$

$$\frac{r \in R \quad P \vdash \Gamma, y : [A]_R}{!x_R(y)_r.P \vdash \Gamma, x : !_r(A)} \ (!\textbf{-pos})$$

$$\frac{r \in R \quad P \vdash \Gamma}{P \vdash \Gamma, x : !_r(A)} \ (!\textbf{-neg-weaken})$$

$$\frac{r \in R \quad P \vdash \Gamma, y : [A]_R}{?x_R[y]_r.P \vdash \Gamma, x : !_r(A)} \ (!\textbf{-neg-derelict})$$

$$\frac{r \in R \quad P \vdash \Gamma, x' : [!_r(A)]_R, x : [!_r(A)]_R}{P\{x/x'\} \vdash \Gamma, x : !_r(A)} \ (!\textbf{-neg-contract})$$

**Figure 5.** The typing rules for $\pi$LMRL

### 3.4 Dynamics of $\pi$LMRL

We present a reduction semantics for $\pi$LMRL based on Lemma 2.6, which states the admissibility of the following rule in LMRL:

$$(\Gamma_1, [A]_{R_1}; \Gamma_2, [A]_{R_2}; \ldots; \Gamma_n, [A]_{R_n}) \Rightarrow \Gamma_1, \Gamma_2, \ldots, \Gamma_n$$

where $\overline{R_1} \uplus \overline{R_2} \uplus \ldots \uplus \overline{R_n} = \overline{\emptyset}$ is assumed. Ideally, we would formulate such a reduction semantics by directly following the proof of Lemma 2.6, which essentially implies following the proof of Lemma2.4. Unfortuately, it is not yet clear to us how the latter can be encoded in a process calculus. Instead, we are to proceed by following a direct proof of Lemma 2.6, which is largely parallel to a standard proof of cut-elimination for classical linear logic (e.g., the one used by Wadler (2012)).

We use $\equiv$ for a structural equivalence relation on processes and assume it to contain the following equivalence rules (perm) and (assoc):

$$\begin{array}{ll} (perm) & \nu x.(\vec{P}_1) \equiv \nu x.(\vec{P}_2) \\ (assoc) & \nu x.(\nu y.(P | \vec{P}_1) | \vec{P}_2) \equiv \nu y.(\nu x.(P | \vec{P}_2) | \vec{P}_1) \end{array}$$

For (perm), $\vec{P}_2$ is assumed to be a permutation of $\vec{P}_1$. For (assoc), both $x \in fn(P_1)$ and $y \in fn(P_1)$ are assumed to hold.

Let us assume a typing derivation $\mathcal{D}$ of the following form (where the last applied rule is (**$n$-cut**) for $n \geq 2$):

$$\frac{\overline{R_1} \uplus \ldots \uplus \overline{R_n} = \overline{\emptyset} \quad \mathcal{D}_i :: P_i \vdash \Gamma_i, x : [A]_{R_i} \text{ for } 1 \leq i \leq n}{\nu x.(P_1 | P_2 | \ldots | P_n) \vdash \Gamma_1, \Gamma_2, \ldots, \Gamma_n}$$

The cut $\nu x.(P_1 | P_2 | \ldots | P_n)$ is a principal cut if each association $x : [A]_{R_i}$ is introduced by the last applied rule in $\mathcal{D}_i$.

We use $\Rightarrow$ for a single-step reduction relation on processes. We first introduce proper rules for reducing principal cuts and then introduce another set of rules for handling non-principal cuts (based on so-called commuting conversions).

**Rules for principal cuts** Let assume that $\nu x.(P_1 | P_2 | \ldots | P_n)$ is principal cut for the moment. Based on the outmost type constructor in $A$, we have four cases of elimination of principal cuts:

- Assume that $A$ is of the form $A_1 \otimes_r A_2$. Without loss of generality, we may assume $r \in \overline{R}_1$, that is, $r \notin R_1$. Then $r \notin \overline{R}_i$ for $2 \le i \le n$, implying $r \in R_i$ for $2 \le i \le n$. It is clear that $P_1$ is of the form $x_{R_1}[y]_r.(P_1^0)$ and $P_i$ of the form $x_{R_i}(y)_r.(P_i^0 | P_i^1)$ for each $2 \le i \le n$. Therefore $\nu x.(P_1 | P_2 | \ldots | P_n)$ can be reduced to the following process $Q_1$:
$$\nu x.(\nu y.(P_1^0 | P_2^0 | \ldots | P_n^0) | P_2^1 | \ldots | P_n^1)$$
which performs a cut of $A_1$ followed by a cut of $A_2$. It can also be reduced to the following process $Q_2$:
$$\nu y.(\nu x.(P_1^0 | P_2^1 | \ldots | P_n^1) | P_2^0 | \ldots | P_n^0)$$
which performs a cut of $A_2$ followed by a cut of $A_1$. Note that $Q_1$ and $Q_2$ are structurally equivalent, that is $Q_1 \equiv Q_2$ holds.
There is a bit of surprise here as $[A_1 \otimes A_2]_R$ (which is $[A_1 \otimes_r A_2]_R$ for $r \in R$) is interpreted as *input* $A_1$ *and then behave as* $A_2$. If it is interpreted as *output* $A_1$ *and then behave as* $A_2$ (which is commonly done in studies on dyadic session types (e.g., (Caires and Pfenning 2010; Wadler 2012a))), then each process $P_i$ needs to send a message to $P_1$ for $2 \le i \le n$, performing a kind of reverse broadcasting.
- Assume that $A$ is of the form $1_r$. Without loss of generality, we may assume $r \notin A_1$ and $r \in A_i$ for $2 \le i \le n$. Clearly, $P_1$ is of the form $x_{R_1}[_]_r.P_1^0$ and $P_i$ of the form $x_{R_i}(_)_r.0$ for each $2 \le i \le n$. Then we have
$$\nu x.(P_1 | P_2 | \ldots | P_n) \Rightarrow P_1^0$$
Note that the last applied rule in $\mathcal{D}_1$ is **(1-neg)**. and the last applied rule in $\mathcal{D}_i$ is **(1-pos)** for $2 \le i \le n$.
- Assume that $A$ is of the form $A_1 \oplus_r A_2$. Without loss of generality, we may assume $r \notin A_1$ and $r \in A_i$ for $2 \le i \le n$. It is clear that $P_1$ is of the form $x_{R_1}[inl]_r.(P_1^0)$ or $x_{R_1}[inr]_r.(P_1^1)$, and $P_i$ is of the form $x_{R_i}(case)_r.(P_i^0 | P_i^1)$. Therefore $\nu x.(P_1 | P_2 | \ldots | P_n)$ can be reduced to the following process
$$\nu x.(P_1^k | P_2^k | \ldots | P_n^k)$$
where $k = 0$ or $k = 1$ (depending on whether *inl* or *inr* occurs in the header of $P_1$).
- Assume that $A$ is of the form $!_r(B)$. Without loss of generality, we may assume $r \notin A_1$ and $r \in A_i$ for $2 \le i \le n$. It is clear that each $P_i$ is of the form $!x_{R_i}(y)_r.(P_i^0)$ for $2 \le i \le n$. As for $P_1$, there are three possibilities.
  - The name $x$ is not in $fn(P_1)$. Then we have
$$\nu x.(P_1 | P_2 | \ldots | P_n) \Rightarrow P_1$$
  - $P_1$ is of the form $?x_{R_1}[y]_r.P_1^0$ for some $P_1^0$ and $x \notin fn(P_1^0)$. Then we can have
$$\nu x.(P_1 | P_2 | \ldots | P_n) \Rightarrow \nu y.(P_1^0 | P_2^0 | \ldots | P_n^0)$$
  - $P_1$ is of the form $P_1^0\{x/x'\}$ for some $P_1^0$ and $x \in fn(P_1^0)$. Then we can have
$$\nu x.(P_1 | P_2 | \ldots | P_n) \Rightarrow \nu x.(\nu x'.(P_1 | P_2' | \ldots | P_n') | P_2 | \ldots | P_n)$$
where $P_i' = !x'_{R_i}(y)_r.(P_i^0)$ for $2 \le i \le n$.

$$
\begin{array}{ll}
\nu x_0.(x[y].P_1 | \vec{P}) \Rightarrow x[y].\nu x_0.(P_1 | \vec{P}) & \\
\nu x_0.(x(y).(P_1 | P_2) | \vec{P}) \Rightarrow x(y).(\nu x_0.(P_1 | \vec{P}) | P_2) & \text{if } x_0 \in fn(P_1) \\
\nu x_0.(x(y).(P_1 | P_2) | \vec{P}) \Rightarrow x(y).(P_1 | \nu x_0.(P_2 | \vec{P})) & \text{if } x_0 \in fn(P_2) \\
\nu x_0.(x[inl].P_1 | \vec{P}) \Rightarrow x[inl].\nu x_0.(P_1 | \vec{P}) & \\
\nu x_0.(x[inr].P_1 | \vec{P}) \Rightarrow x[inr].\nu x_0.(P_1 | \vec{P}) & \\
\nu x_0.(x(case).(P_1, P_2) | \vec{P}) \Rightarrow x(case).\nu x_0.(P_1 | \vec{P}, P_2 | \vec{P}) & \\
\nu x_0.(!x[y].P_1 | \vec{P}) \Rightarrow !x[y].\nu x_0.(P_1 | \vec{P}) & \\
\nu x_0.(?x[y].P_1 | \vec{P}) \Rightarrow ?x[y].\nu x_0.(P_1 | \vec{P}) &
\end{array}
$$

**Figure 6.** The rules for commuting conversions

$$\frac{r \in R \wedge s \in R \quad P \vdash \Gamma, x : [A]_R}{x_R(skip)_{r,s}.P \vdash \Gamma, x : [msg_{r,s}(A)]_R}\ \textbf{(msg-pos-pos)}$$

$$\frac{r \notin R \wedge s \notin R \quad P \vdash \Gamma, x : [A]_R}{x_R[skip]_{r,s}.P \vdash \Gamma, x : [msg_{r,s}(A)]_R}\ \textbf{(msg-neg-neg)}$$

$$\frac{r \in R \wedge s \notin R \quad P \vdash \Gamma, x : [A]_R}{x_R(recv)_{r,s}.P \vdash \Gamma, x : [msg_{r,s}(A)]_R}\ \textbf{(msg-pos-neg)}$$

$$\frac{r \notin R \wedge s \in R \quad P \vdash \Gamma, x : [A]_R}{x_R[send]_{r,s}.P \vdash \Gamma, x : [msg_{r,s}(A)]_R}\ \textbf{(msg-neg-pos)}$$

**Figure 7.** Typing rules for $msg_{r,s}$

We have covered all of the possible cases for $P_1$.

Please find in Figure 8 some detailed illustration of the reduction rules for eliminating principal cuts.

**Rules for non-principal cuts** The rules for handling non-principal cuts are based on commuting conversions, each of which pushes a cut inside a communication header. If can be readily checked that a cut must be a principal one of none of the rules in Figure 6 are applicable. Please see (Wadler 2014) for details on commuting conversions.

### 3.5 Subject Reduction and Cut-Elimination

Let us define $P \Longrightarrow Q$ as $P \equiv P_1$ and $P_1 \Rightarrow Q_1$ and $Q_1 \equiv Q$ for some $P_1$ and $Q_1$. Let $\Longrightarrow^*$ be the transitive closure of $\Longrightarrow$.

**Theorem 3.1** (Subject Reduction)**.** *Assume that $P \vdash \Gamma$ is derivable. If $P \Longrightarrow P'$, then $P' \vdash \Gamma$ is also derivable.*

*Proof.* It simply follows the way in which $\Longrightarrow$ is defined. □

**Theorem 3.2** (Cut-Elimination)**.** *Assume that $P \vdash \Gamma$ is derivable. If $P$ is a cut, then $P \Longrightarrow P'$ holds for some $P'$.*

*Proof.* (Sketch) It follows from a simple case analysis on $P$. For details, one may see the proof of Theorem 2 (Wadler 2014). Note that we use cut-elimination in a rather liberal sense: It means the elimination of a particular cut (and $P'$ may be allowed to contain cuts). The point is to guarantee progress being made (rather than the eventuality of reaching some sort of terminal state). □

### 3.6 Support for Point-to-Point Communication

By inspecting the reduction rules in $\pi$LMRL that involve communication, we can clearly see that only communication in the form of broadcasting is involved. However, it is rather simple to support point-to-point communication by extending $\pi$LMRL with unary type constructors $msg_{r,s}$, where $r$ and $s$ range over all of the distinct pairs of roles.

$$\nu x.(x[skip].P_1 \,|\, x(skip).P_2 \,|\, \ldots \,|\, x(skip).P_n)$$
$$\Rightarrow \nu x.(P_1 \,|\, P_2 \,|\, \ldots \,|\, P_n)$$
$$\nu x.(x[send].P_1 \,|\, x(recv).P_2 \,|\, x(skip).P_3 \,|\, \ldots \,|\, x(skip).P_n)$$
$$\Rightarrow \nu x.(P_1 \,|\, P_2 \,|\, P_3 \,|\, \ldots \,|\, P_n)$$

## 4. Related Work and Conclusion

Session types were introduced by Honda (Honda 1993) and further extended subsequently (Takeuchi et al. 1994; Honda et al. 1998). There have since been extensive theoretical studies on session types in the literature(e.g., (Castagna et al. 2009; Gay and Vasconcelos 2010; Caires and Pfenning 2010; Toninho et al. 2011b; Vasconcelos 2012; Wadler 2012a; Lindley and Morris 2015)). Multiparty session types, as a generalization of (dyadic) session types, were introduced by Honda and others (Honda et al. 2008), together with the notion of global types, local types, projection and coherence.

Introduced by Milner and others (Milner et al. 1992a,b), $\pi$-calculus allows channel names to be communicated along the channels themselves, making it possible to describe concurrent computations with changing network configuration. Connections between $\pi$-calculus and linear logic have been actively studied (Abramsky 1994b; Bellin and Scott 1994a) from early on, and it is demonstrated in some recent work (Caires and Pfenning 2010; Wadler 2012b, 2014) that a tight proofs-as-processes correspondences exists for dyadic sessions. And some of closely related additional work includes Toninho et al. (2011a); Pfenning et al. (2011); Toninho et al. (2012); Pérez et al. (2012).

Continuing this line of works, Carbone (Carbone et al. 2015) introduced MCP, a variant of CLL that admits MCut, a generalized cut-rule for composing multiple proofs. MCut requires coherent proofs (obtained through a separate proof system) as a side condition. This work is probably the first along the line that interprets $\otimes$ as input and $⅋$ as output (as opposed to all of the other works we are aware of). Their follow-up work (Carbone et al. 2016) introduced a variant of MCP, and a translation from MCP to CP (Wadler 2012b) via GCP (some intermediate calculus) that interprets a coherence proof as an arbiter process that mediates communications in a multiparty session. But it reverts to $\otimes$ as output and $⅋$ as input. In this paper, we take a very different route to formulate a cut-rule for multiple sequents, naturally extending the celebrated result of cut-elimination by Gentzen.

There have been studies on multirole parties (Yoshida and Deniélou 2011; Neykova and Yoshida 2014), where such parties play multiple roles by holding channels belonging to multiple sessions. We see no direct relation between a multirole party and a multirole channel as is formulated in this paper.

There is also very recent work on encoding multiparty session types based on binary session types (Caires and Pérez 2016), which relies on an arbiter process to mediate communications between multiple parties while preserving global sequencing information. Clearly, this form of mediating (formulated based on automata theory) is closely related to performing a cut to multiple processes in $\pi$LMRL.

While multirole logic stems from studies on multiparty session types, it is certainly not restricted to such studies. Just as the notion of linearity (as in linear logic) that has greatly enriched the study on logics and programming languages, we hope that the notion of multirole (as in multirole logic) can exert a significant impact in this regard as well.

$$\dfrac{\dfrac{}{id(x_1,x)\vdash x_1,x}\text{Axiom}\quad \dfrac{}{id(x_2,x)\vdash x_2,x}\text{Axiom}\quad \dfrac{}{id(x_3,x)\vdash x_3,x}\text{Axiom}}{\nu x.(id(x_1,x)\,|\,id(x_2,x)\,|\,id(x_3,x))\vdash x_1,x_2,x_3}\text{3-Cut}$$

$$\Longrightarrow\quad \dfrac{}{id(x_1,x_2,x_3)\vdash x_1,x_2,x_3}\text{Axiom}\qquad(\beta_{Ax})$$

$$\dfrac{\dfrac{r\notin R_1\quad P\vdash y,x}{x[y].P\vdash x}\otimes\text{-neg}\quad \dfrac{r\in R_2\quad P_2\vdash y_2\quad Q_2\vdash x}{x(y_2).(P_2|Q_2)\vdash x}\otimes\text{-pos}\quad \dfrac{r\in R_3\quad P_3\vdash y_3\quad Q_3\vdash x}{x(y_3).(P_3|Q_3)\vdash x}\otimes\text{-pos}}{\nu x.(x[y].P\,|\,x(y_2).(P_2\,|\,Q_2)\,|\,x(y_3).(P_3\,|\,Q_3))\vdash\cdot}\text{3-Cut}$$

$$\Longrightarrow\quad \dfrac{\dfrac{P\vdash y,x\quad Q_2\vdash x\quad Q_3\vdash x}{\nu x.(P\,|\,Q_2\,|\,Q_3)\vdash y}\text{3-Cut}\quad P_2\{y/y_2\}\vdash y\quad P_3\{y/y_3\}\vdash y}{\nu y.(\nu x.(P\,|\,Q_2\,|\,Q_3)\,|\,P_2\{y/y_2\}\,|\,P_3\{y/y_3\})\vdash\cdot}\text{3-Cut}\qquad(\beta_{\otimes})$$

$$\dfrac{\dfrac{r\notin R_1\quad P\vdash x}{x[inl].P\vdash x}\&\text{-neg-l}\quad \dfrac{r\in R_2\quad P_2\vdash x\quad Q_2\vdash x}{x.case(P_2,Q_2)\vdash x}\&\text{-pos}\quad \dfrac{r\in R_3\quad P_3\vdash x\quad Q_3\vdash x}{x.case(P_3,Q_3)\vdash x}\&\text{-pos}}{\nu x.(x[inl].P\,|\,x.case(P_2,Q_2)\,|\,x.case(P_3,Q_3))\vdash\cdot}\text{3-Cut}$$

$$\Longrightarrow\quad \dfrac{P\vdash x\quad P_2\vdash x\quad P_3\vdash x}{\nu x.(P|P_2|P_3)\vdash\cdot}\text{3-Cut}\qquad(\beta_{\&l})$$

$$\dfrac{\dfrac{r\notin R_1\quad P\vdash x}{x[inr].P\vdash x}\&\text{-neg-r}\quad \dfrac{r\in R_2\quad P_2\vdash x\quad Q_2\vdash x}{x.case(P_2,Q_2)\vdash x}\&\text{-pos}\quad \dfrac{r\in R_3\quad P_3\vdash x\quad Q_3\vdash x}{x.case(P_3,Q_3)\vdash x}\&\text{-pos}}{\nu x.(x[inr].P\,|\,x.case(P_2,Q_2)\,|\,x.case(P_3,Q_3))\vdash\cdot}\text{3-Cut}$$

$$\Longrightarrow\quad \dfrac{P\vdash x\quad Q_2\vdash x\quad Q_3\vdash x}{\nu x.(P|Q_2|Q_3)\vdash\cdot}\text{3-Cut}\qquad(\beta_{\&r})$$

$$\dfrac{\dfrac{r\notin R_1\quad P\vdash\cdot}{P\vdash x}\mathbf{1}\text{-neg}\quad \dfrac{r\in R_2}{\mathbf{0}\vdash x}\mathbf{1}\text{-pos}\quad \dfrac{r\in R_3}{\mathbf{0}\vdash x}\mathbf{1}\text{-pos}}{\nu x.(P\,|\,\mathbf{0}\,|\,\mathbf{0})\vdash\cdot}\text{3-Cut}\quad\Longrightarrow\quad P\vdash\cdot\qquad(\beta_{\mathbf{1}})$$

$$\dfrac{\dfrac{r\notin R_1\quad P\vdash y}{?x(y).P\vdash x}\text{!-neg-derelict}\quad \dfrac{r\in R_2\quad P_2\vdash y_2}{!x(y_2).P_2\vdash x}\text{!-pos}\quad \dfrac{r\in R_3\quad P_3\vdash y_3}{!x(y_3).P_3\vdash x}\text{!-pos}}{\nu x.(?x[y].P\,|\,!x(y_2).P_2\,|\,!x(y_3).P_3)\vdash\cdot}\text{3-Cut}$$

$$\Longrightarrow\quad \dfrac{P\vdash y\quad P_2\{y/y_2\}\vdash y\quad P_3\{y/y_3\}\vdash y}{\nu y.(P\,|\,P_2\{y/y_2\}\,|\,P_3\{y/y_3\})\vdash\cdot}\text{3-Cut}\qquad(\beta_{!})$$

$$\dfrac{\dfrac{r\notin R_1\quad P\vdash\cdot}{P\vdash x}\text{!-neg-weaken}\quad \dfrac{r\in R_2\quad P_2\vdash y_2}{!x(y_2).P_2\vdash x}\text{!-pos}\quad \dfrac{r\in R_3\quad P_3\vdash y_3}{!x(y_3).P_3\vdash x}\text{!-pos}}{\nu x.(P\,|\,!x(y_2).P_2\,|\,!x(y_3).P_3)\vdash\cdot}\text{3-Cut}\quad\Longrightarrow\quad \dfrac{P\vdash\cdot}{P\vdash\cdot}\text{!-neg-weaken}\qquad(\beta_{!W})$$

$$\dfrac{\dfrac{r\notin R_1\quad P\vdash m,n}{P\{x/m,n\}\vdash x}\text{!-neg-contract}\quad \dfrac{r\in R_2\quad P_2\vdash y_2}{!x(y_2).P_2\vdash x}\text{!-pos}\quad \dfrac{r\in R_3\quad P_3\vdash y_3}{!x(y_3).P_3\vdash x}\text{!-pos}}{\nu x.(P\{x/m,n\}\,|\,!x(y_2).P_2\,|\,!x(y_3).P_3)\vdash\cdot}\text{3-Cut}$$

$$\Longrightarrow\quad \dfrac{\dfrac{\dfrac{P\vdash m,n\quad \dfrac{r\in R_2\quad P_2'\vdash y_2}{!m(y_2).P_2'\vdash m}\text{!-pos}\quad \dfrac{r\in R_3\quad P_3'\vdash y_3}{!m(y_3).P_3'\vdash m}\text{!-pos}}{\nu m.(P\,|\,!m(y_2).P_2'\,|\,!m(y_3).P_3')\vdash n}\text{3-Cut}\quad \dfrac{r\in R_2\quad P_2''\vdash y_2}{!n(y_2).P_2''\vdash n}\text{!-pos}\quad \dfrac{r\in R_3\quad P_3''\vdash y_3}{!n(y_3).P_3''\vdash n}\text{!-pos}}{\nu n.(\nu m.(P\,|\,!m(y_2).P_2'\,|\,!m(y_3).P_3')\,|\,!n(y_2).P_2''\,|\,!n(y_3).P_3'')\vdash\cdot}\text{3-Cut}}{\nu n.(\nu m.(P\,|\,!m(y_2).P_2\,|\,!m(y_3).P_3)\,|\,!n(y_2).P_2\,|\,!n(y_3).P_3)\vdash\cdot}\text{!-neg-contract}\qquad(\beta_{!C})$$

**Figure 8.** Principal Cut Reduction for 3-Party Sessions

Without loss of generality, we present here a 3-party principal cut reduction without proof terms. Assume $\overline{R}_1\uplus\overline{R}_2\uplus\overline{R}_3=\overline{\varnothing}$ and $r\notin R_1$